\documentclass[prl,twocolumn,showpacs]{revtex4}
\usepackage{graphicx}

\begin{document}

\title{Generalized diffusion equation}
\author{Jean Pierre Boon}
\email{jpboon@ulb.ac.be}
\homepage{http://poseidon.ulb.ac.be/}
\author{James F. Lutsko}
\email{lutsko@ulb.ac.be}
\affiliation{Center for Nonlinear Phenomena and Complex Systems\\
Universit\'{e} Libre de Bruxelles, 1050 - Bruxelles, Belgium}
\date{\today}

\begin{abstract}
Modern analyses of diffusion processes have proposed nonlinear versions of
the Fokker-Planck equation to account for non-classical diffusion. These
nonlinear equations are usually constructed on a phenomenological basis.
Here we introduce a nonlinear transformation by defining the $q$-generating
function which, when applied to the intermediate scattering function of
classical statistical mechanics, yields, in a mathematically systematic
derivation, a generalized form of the advection-diffusion equation in
Fourier space. Its solutions are discussed and suggest that the $q$%
-generating function approach should be a useful tool to generalize
classical diffusive transport formulations.
\end{abstract}

\pacs{05.90.+m, 05.60.-k, 05.10.Gg}
\maketitle

The development of generalizations of classical statistical mechanical
paradigms to describe non-equilibrium systems has become an important area
of activity. One of the best known examples of this approach is the
so-called ''nonextensive'' thermodynamics first introduced by Tsallis \cite%
{tsallis} which is based on a new definition of entropy that as being the $q$%
-logarithm of the number of states, rather than the natural logarithm in
Boltzmann's definition. It has been shown that thermodynamics can be
consistantly built using the Tsallis entropy as a starting point \cite%
{gellmann-tsallis}. The success of this program has led to the introduction
of other ''nonextensive '' generalizations such as the nonextensive
diffusion equation \cite{swinney-tsallis}. As in the case of the entropy,
these generalizations are usually based on the replacement of a classical,
extensive (frequently linear) phenomenological expression by a
parameterized, nonextensive (frequently non-linear) expression which reduces
to the classical expression for some value of the parameter. In this paper,
we show that for the particular case of diffusion it is possible to
introduce these generalizations at a more fundamental level. The key point
is that statistical mechanical derivations often make use of resummations
and ansatzs which can be made ''nonextensive '' in a natural way since they
are intrinsically heuristic.

Processes related to diffusive transport are most commonly measured by
scattering methods (photon correlation and thermal neutron spectroscopy) and
the spectra are analyzed using the intermediate scattering function~\cite%
{boon-yip} 
\begin{equation}
F_{s}(k,t)=\langle \exp [\imath \mathbf{k}\cdot (\mathbf{R}(t)-\mathbf{R}%
(0))]\rangle \,,  \label{scat_fct}
\end{equation}%
which is the space-Fourier transform of the single particle density
correlation function, i.e. the probability that a tagged particle be at
position ${\mathbf{R}(t)}$ at time $t$, given it was at ${\mathbf{R}(0)}$ at
the initial time. In classical statistical mechanics, one shows that this
function obeys the usual diffusion - or advection-diffusion - equation in
the hydrodynamic limit, i.e. for long wavelengths: $k\ell <<1$, where $\ell $
is the mean free path. This is easily obtained, for instance, by the
cumulant expansion of $F_{s}(k,t)$ \cite{boon-yip}. Here we generalize the
procedure by defining the $q$-generating function 
\begin{equation}
\mathcal{G}(k,t)\equiv \ln _{q}F_{s}(k,t)=\sum_{j=0}^{\infty }\frac{(\imath 
\mathbf{k})^{j}}{j!}\cdot \mathsf{K}_{j}(t)\,,  \label{gen_fct}
\end{equation}%
or equivalently the $q$-cumulant expansion 
\begin{eqnarray}
F_{s}(k,t) &=&\exp _{q}\left[ \sum_{j=0}^{\infty }\frac{(\imath \mathbf{k}%
)^{j}}{j!}\cdot \mathsf{K}_{j}(t)\right]  \nonumber \\
&=&\left[ 1+(1-q)\sum_{j=0}^{\infty }\frac{(\imath \mathbf{k})^{j}}{j!}\cdot 
\mathsf{K}_{j}(t)\right] ^{\frac{1}{1-q}}\;,  \label{cum_exp}
\end{eqnarray}%
where $\mathsf{K}_{j}(t)$ are time dependent functions which can be
expressed in terms of the spatial moments \cite{boon-yip}. Note that in the
limit $q\rightarrow 1$, the $q$-log and the $q$-exp reduce to the usual
logarithmic and exponential expressions \cite{tsallis}. From (\ref{cum_exp}%
), we have 
\begin{equation}
\frac{\partial {F_{s}(k,t)}}{\partial {t}}\,=\,\sum_{j=0}^{\infty }\frac{%
(\imath \mathbf{k})^{j}}{j!}\cdot \left[ \frac{\partial \mathsf{K}_{j}(t)}{%
\partial {t}}\right] F_{s}^{q}(k,t)\,,  \label{t_deriv}
\end{equation}%
which is an exact equation that follows from the nonlinear transformation (%
\ref{gen_fct}). Low $k$ expansion on the r.h.s. of (\ref{t_deriv}) yields 
\begin{equation}
\frac{\partial {F_{s}(k,t)}}{\partial {t}}\,=\,\left[ \imath \mathbf{k}\cdot 
\frac{\partial {{\mbox{\boldmath $\kappa$}}_{1}(t)}}{\partial {t}}-\frac{1}{2%
}k^{2}\frac{\partial {\kappa _{2}(t)}}{\partial {t}}\,+\,...\right]
F_{s}^{q}(k,t)\,,  \label{F_expan}
\end{equation}%
where 
\begin{equation}
\frac{\partial {{\mbox{\boldmath $\kappa$}}}_{1}(t)}{\partial {t}}\,=\,\frac{%
\partial {}}{\partial {t}}\langle \lbrack \mathbf{R}(t)-\mathbf{R}%
(0)]\rangle \,=\,\mathbf{u}+\langle \mathbf{v}(t)\rangle \,=\,\mathbf{u}\,,
\label{kappa_1}
\end{equation}%
with $\mathbf{u}$ the macroscopic velocity or ''drift term '', and 
\begin{equation}
\frac{\partial {\kappa _{2}(t)}}{\partial {t}}\,=\,\frac{\partial {}}{%
\partial {t}}\langle \lbrack \mathbf{R}(t)-\mathbf{R}(0)]^{2}\rangle
\,=\,2\,D\,,  \label{kappa_2}
\end{equation}%
where, for simplicity, we have considered isotropic diffusion, with $D$ the
self-diffusion coefficient. So to second order, as usually considered for
the classical hydrodynamic limit, we obtain the nonlinear
advection-diffusion equation 
\begin{equation}
\frac{\partial {F_{s}(k,t)}}{\partial {t}}\,=\,\imath \mathbf{k}\cdot 
\mathbf{u}\,{F_{s}^{q}(k,t)}\,-\,Dk^{2}\,{F_{s}^{q}(k,t)}\,,
\label{q-equation}
\end{equation}%
or 
\begin{equation}
\frac{1}{1-q}\frac{\partial {F_{s}^{1-q}(k,t)}}{\partial {t}}\,\,=\,\imath 
\mathbf{k}\cdot \mathbf{u}-Dk^{2}\,.
\end{equation}%
The solution for initial condition $F_{s}(k,t=0)=1$ (i.e. a real space
delta-function) and for positive times is 
\begin{eqnarray}
F_{s}(k,t)\,&=&\,\left[ 1+(1-q)(\imath \mathbf{k}\cdot \mathbf{u}-Dk^{2})t%
\right]^{\frac{1}{1-q}}  \nonumber \\
\,&\equiv &\exp _{q}\left[(\imath \mathbf{k}\cdot \mathbf{u}-Dk^{2})t\right]%
\;,  \label{sol_q-eq}
\end{eqnarray}%
a form commonly found in nonextensive statistical mechanics \cite%
{tsallis-swinney}. It is clear that, in the limit $q~\rightarrow~1$, one
retrieves the classical result: $F_{s}(k,t)\,=\,e^{(\imath \mathbf{k}\cdot 
\mathbf{u}-Dk^{2})t}$.

Equation (\ref{q-equation}) rewritten as 
\begin{equation}
\frac{\partial {F_{s}(k,t)}}{\partial {t}}\,-\,\imath \mathbf{k}\cdot 
\mathbf{u}_q(k,t)\,{F_{s}(k,t)}\,=\,-\,Dk^{2}\,{F_{s}^{q}(k,t)}\,,
\label{pm-equation}
\end{equation}%
with $\mathbf{u}_q(k,t)\,=\,\mathbf{u}\,F_{s}^{q-1}(k,t)$, can be viewed as
the analogue in Fourier space of the porous media equation \cite%
{tsallis-buckman,abe}, but in contrast to the latter, the nonlinear
advection-diffusion equation (\ref{pm-equation}) yields generalized, but
classical diffusive behavior in real space (i.e. with $r/\sqrt{Dt}$ scaling,
as we shall see below), and therefore applies to a different class of
diffusive transport.

In the absence of flow ($\mathbf{u}=0$), we call Eq.(\ref{q-equation}) the 
\emph{q-diffusion equation} 
\begin{equation}
\frac{\partial {F_{s}(k,t)}}{\partial {t}}\,=\,-Dk^{2}F_{s}^{q}(k,t)\,.
\label{q-diff_eq}
\end{equation}%
This generalized equation can be cast into a regular diffusion equation with
an effective diffusion coefficient $D_{eff}$ by using the identification 
\begin{equation}
D_{eff}\equiv D_{q}\,=\,D\,F_{s}^{q-1}(k,t)\,,  \label{diff_eff}
\end{equation}%
or, substituting the solution given in (\ref{sol_q-eq}) (with $\mathbf{u}=0$%
) and expanding in $Dk^2t$ (i.e. for long wavelengths), 
\begin{eqnarray}
D_{q} \,=\, D\,\sum_{j=0}^{\infty } {(1-q)^{j}}\,(Dk^{2}t)^{j} \,, 
\label{D_sum}
\end{eqnarray}%
that is $D_{q}\,=\, D\,+\,higher\;moments\,(q,k,t)$, where the higher
moments depend on $q$, and also on $k$ and $t$ as for the space- and
time-dependent transport coefficients in generalized hydrodynamics \cite%
{boon-yip}. For $q=1$, only the term $j=0$ survives in the sum in (\ref%
{D_sum}), and one retrieves the classical diffusion formulation.

We now consider what type of distributions are obtained from the 
solution to the $q$-diffusion equation (\ref{q-diff_eq}): 
\begin{equation}
F_{s}(k,t)=\left[ 1-\left( 1-q\right) Dk^{2}t\right] ^{\frac{1}{1-q}}\,.
\label{q-sol}
\end{equation}%
As the scattering function must be real and non-negative, the argument in
brackets must be positive. This is always realized for $q>1$. If we are to
allow $q<1$, then it is necessary to limit the support of the intermediate
scattering function by including a step function giving%
\begin{equation}
F_{s}(k,t)=\left[ 1-\left( 1-q\right) Dk^{2}t\right] ^{\frac{1}{1-q}}\Theta
\left( 1-\left( 1-q\right) Dk^{2}t\right) \,,  \label{sol_theta}
\end{equation}%
where $\Theta \left( y \right) =1$ for $y >1$ and $\Theta \left( y \right) =0
$ otherwise. Substitution of (\ref{sol_theta}) into Eq.(\ref{q-diff_eq})
shows that (\ref{sol_theta}) satisfies the $q$-diffusion equation since the
time-derivative acting on the step function gives a Dirac delta function
with a prefactor that goes to zero. Note that in this case, since the
intermediate scattering function is a function of $Dk^{2}t$, it immediately
follows that in real space it is a function of $r^{2}/Dt$ so that $%
\left\langle r^{2}\right\rangle \sim Dt$ which corresponds to normal
diffusion, albeit with non-classical distribution functions.

In one dimension, inversion of the Fourier transform (\ref{q-sol}) in the
case $q>1$ gives%
\begin{equation}
G_{q>1}(r,t)=\frac{\Gamma ^{-1}\left( \frac{1}{q-1}\right) }{\sqrt{\pi }%
\sqrt{\left( q-1\right) Dt}}\left( \frac{x}{2}\right) ^{\frac{1}{q-1}-\frac{1%
}{2}}K_{\frac{1}{q-1}-\frac{1}{2}}(x)\,,  \label{gen_sol_q>1}
\end{equation}%
where $x=\frac{r}{\sqrt{\left( q-1\right) Dt}}$, and 
$K_{\nu }\left( x\right) $ is the modified Bessel function of the second
kind. The structure of this result should not be too surprising since we
know that the general solution to the classical diffusion problem is given
in terms of Bessel functions \cite{arfken}. For some rational values of $q$
it is possible to express the Bessel function in terms of elementary
functions in which case the solution assumes the form $p(x)\exp (-x)$ where $%
p(x)$ is a polynomial in $x$. For instance for $q=3/2$, we obtain 
\begin{equation}
G_{3/2}(r,t)\,=\,\frac{1}{2\sqrt{2Dt}}\left( x+1\right) \exp \left(
-x\right) \,.  \label{sol_3/2}
\end{equation}%
Figure 1 shows a numerical evaluation of the solutions for various values of 
$q$, and, as $q$ increases, it does indeed appear to decay as an
exponential. 

We now show that (\ref{gen_sol_q>1}) does give the correct Gaussian limit
as $q \rightarrow 1$. We make explicit the dependence on $q$ by writing%
\begin{equation}
K_{\frac{1}{q-1}-\frac{1}{2}}(x)=K_{\frac{1}{q-1}-\frac{1}{2}}\left( \frac{r^{\ast}%
}{\sqrt{q-1}}\right) \,,
\end{equation}%
where $r^{\ast}\,=\,{r}/\sqrt{\mathit{Dt}}$. Introducing $\nu =\frac{1}{q-1}-\frac{1}{2}$ 
in (\ref{gen_sol_q>1})  gives%
\begin{eqnarray}
G_{q>1}(r,t)&=&\frac{\Gamma ^{-1}\left( \nu +\frac{1}{2}\right) }{\sqrt{\pi }%
\sqrt{Dt}}\sqrt{\nu +\frac{1}{2}}\left( \frac{r^{\ast}}{2}\sqrt{\nu +\frac{1}{2}}%
\right) ^{\nu }\times \nonumber \\
&\times& K_{\nu }\left( r^{\ast} \sqrt{\nu +\frac{1}{2}}\right) \,.
\end{eqnarray}%
Using the asymptotic expansion for $\nu \gg 1$ \cite{abramowitz}
\begin{eqnarray}
K_{\nu }(\nu z)&\sim& \sqrt{\frac{\pi }{2\nu }}\left( \frac{z}{1+\sqrt{1+z^{2}}%
}\right) ^{-\nu }\frac{1}{\left( 1+z^{2}\right) ^{1/4}} \times \nonumber \\
&\times& \exp \left( -\nu \left( \sqrt{1+z^{2}}\right) \right) \,, \nonumber
\end{eqnarray}%
with $z=r^{\ast} \frac{\sqrt{\nu +\frac{1}{2}}}{\nu }$,  gives, for fixed $r^{\ast}$,%
\begin{eqnarray}
K_{\nu}(r^{\ast} {\sqrt{\nu +\frac{1}{2}}}) \sim \sqrt{\frac{\pi }{2\nu }}\left( r^{\ast}%
\frac{\sqrt{\nu +\frac{1}{2}}}{\nu }\right) ^{-\nu } \times \nonumber \\
\times \exp \left( \nu \allowbreak \ln 2+\frac{1}{4}{r^{\ast}}^{2}\right)
\exp \left( -\nu -\frac{1}{2} {r^{\ast}}^{2}\right) \nonumber \\
\sim \sqrt{\frac{\pi }{2\nu }}\left({r^{\ast}}  \frac{\sqrt{\nu +\frac{1}{2}}}{2\nu }%
\right) ^{-\nu }\exp \left( -\nu -\frac{1}{4}{r^{\ast}}^{2}\right) \,,  \nonumber
\end{eqnarray}%
wherefrom we obtain%
\begin{equation}
G_{q>1}(r,t)\sim \frac{\Gamma ^{-1}\left( \nu +\frac{1}{2}\right) }{\sqrt{2Dt%
}}\;\nu ^{\nu }\exp \left( -\nu -\frac{1}{4}{r^{\ast}}^{2}\right) \,,
\end{equation}%
and, using Stirling's formula, we find%
\begin{equation}
G_{q>1}(r,t)\sim \frac{1}{2\sqrt{\pi Dt}}\exp \left( -\frac{r^{2}}{4Dt} \right)\,,%
\end{equation}%
as expected. Fig.2 clearly shows that the distribution approaches a Gaussian
in the limit $\nu \gg 1$ or $q \rightarrow 1$.

For $q<1$, Fourier inversion of (\ref{q-sol}) yields%
\begin{equation}
G_{q<1}(r,t)=\frac{1}{2 \sqrt{\pi }}\frac{\Gamma \left( \frac{1}{1-q}+1\right)%
}{\sqrt{\left( 1-q\right) Dt}}\, \left( \frac{x}{2}\right) ^{-\frac{1}{1-q}-%
\frac{1}{2}}{J}_{\frac{1}{1-q}+\frac{1}{2}}\left( x\right) \,, \allowbreak
\label{gen_sol_q<1}
\end{equation}%
with $x=\frac{r}{\sqrt{\left( 1-q \right) Dt}}$, and
where ${J}_{\nu}$ is the Bessel function of the first kind. Here also, for
some values of $q$, this result can be written in terms of elementary
functions in the form $l(x)\cos x+m(x)\sin x$ where $l(x)$ and $m(x)$ are
polynomials in $x$. As an example, for $q = 1/2$, we have 
\begin{equation}
G_{1/2}(r,t) = \frac{{\pi}^{-1}}{\sqrt{2Dt}}\left( \frac{x}{2} \right)^{-5}%
\left(3 \sin x - 3x \cos x - x^{2} \sin x \right) .  \label{sol_1/2}
\end{equation}%
Figure 2 shows that as $q \rightarrow 1$ the distribution approaches a
Gaussian, and Fig. 3 shows $G_{q}(r,t)$ for several values of $q<1$. In this
case, we observe that the distribution goes sometimes negative (for small
values of $q$) and, hence, becomes unphysical. Note that after the
distribution first becomes negative, its magnitude is always very small so
that in some sense it appears to be approximating a function with finite
support. Therefore, we must either (a) restrict the range of allowable
values of $q$ to $q\geq 1$, or (b) allow values of $q<1$ with the
understanding that the distribution is an approximation to a
positive-definite function, perhaps one having finite support. It is worth
noting in passing that option (b) is quite commonly used in other areas of
statistical mechancs. For example, the Chapman-Enskog solution to the
Boltzmann equation results, at all orders, in a distribution which is not
positive-definite \cite{chapman}.

It is interesting to ask what conventional diffusion process would yield the
above results. Casting the general solutions (\ref{gen_sol_q>1}) and (\ref%
{gen_sol_q<1}) into the form of a  generalized diffusion equation in real space 
\[
\frac{\partial {G_{q}(r,t)}}{\partial {t}}\,=\,D_{q}(r,t)\,\frac{\partial
^{2}{G_{q}(r,t)}}{\partial {r^{2}}}\;.
\]%
gives, for $q>1$, 
\begin{equation}
D_{q>1}(r,t)\,=\,D\,\frac{q-1}{2}\,\frac{N(r,t)}{M(r,t)}\,,  \label{Dq>1}
\end{equation}%
where, with the notation $\nu ={\frac{1}{q-1}}\,-\frac{1}{2}$, 
\begin{eqnarray}
N(r,t)&\,=\,&\frac{r}{\sqrt{{(q-1)Dt}}}\,\frac{\mathit{K}_{\nu +1}\left( {%
\frac{r}{\sqrt{{(q-1)Dt}}}}\right) }{\mathit{K}_{\nu }\left( {\frac{r}{\sqrt{%
{(q-1)Dt}}}}\right) }-1-2\,\nu\,, \nonumber \\
M(r,t)&\,=\,&1+\frac{(q-1)Dt}{r^{2}}\,(1-2\nu )\left( N(r,t)+1\right) \,, \nonumber
\end{eqnarray}%
and, for $q<1$, 
\begin{equation}
D_{q<1}(r,t)\,=\,D\frac{1-q}{2}\,\frac{N^{\prime }(r,t)}{M^{\prime }(r,t)}\,,
\label{Dq<1}
\end{equation}%
where, with the notation $\nu ={\frac{1}{1-q}}\,+\frac{1}{2}$, 
\begin{eqnarray}
N^{\prime }(r,t) &\,=\,&\frac{r}{\sqrt{{(1-q)Dt}}}\,\frac{\mathit{J}_{\nu
+1}\left( {\frac{r}{\sqrt{{(1-q)Dt}}}}\right) }{\mathit{J}_{\nu }\left( {%
\frac{r}{\sqrt{{(1-q)Dt}}}}\right) }-1\,,  \nonumber  \label{NM<} \\
M^{\prime }(r,t) &\,=\,&-1+\frac{{(1-q)Dt}}{r^{2}}(1+2\nu )\left( N^{\prime
}(r,t)+1\right) \,. \nonumber
\end{eqnarray}

Two typical examples (for $q=3/2$ and $q=1/2$) of the generalized diffusion
coefficient are \cite{footnote}
\begin{equation}
D_{3/2}(r,t)\,=\,\frac{D^{\prime }}{2}\,\left( \frac{r}{\sqrt{{D^{\prime }t}}%
}+\frac{1}{1-\frac{r}{\sqrt{{D^{\prime }t}}}}\right) \,,  \label{D_3/2}
\end{equation}%
where ${D^{\prime }}=\frac{1}{2}D$, and
\begin{equation}
D_{1/2}(r,t)\,=\,\frac{D^{\prime }}{2}\,\frac{\ {x}^{2}\sin\,x+(3-x)(x\cos
\,x-\sin \,x)}{4\,\sin \,x+(\frac{12}{x}-1)(\cos \,x-\frac{\sin \,x}{x})}\,,
\label{D_1/2}
\end{equation}%
with  $x=\frac{r}{\sqrt{{D^{\prime }t}}}\,$. 

Although the diffusion coefficient $D_{q}$ exhibits a generalized form
because of its dependence on the distribution function $G_{q}(r,t)$,
diffusion remains normal in the sense that $\frac{r}{\sqrt{{Dt}}}$ scaling
is preserved as we already observed and as can be seen even more explicitly
by computing the mean square displacement 
\begin{equation}
\langle r^{2}(t)\rangle \,=\,\int_{0}^{\infty }dr\;r^{2}G_{q}(r,t)\,,
\label{r2t}
\end{equation}%
with $G_{q}(r,t)$ given either by (\ref{gen_sol_q>1}) or (\ref{gen_sol_q<1}%
), which, with $r^{\ast }\,=\,{r}/\sqrt{\mathit{Dt}}$, yields 
\begin{equation}
\langle r^{2}(t)\rangle \,=\,\,{Dt}\,2\int_{0}^{\infty }dr^{\ast }\;{r^{\ast }}%
^{2}\,G_{q}(r^{\ast })\,=\,2\,D\,t\;.  \label{rDt>}
\end{equation}

In this paper, we have shown how a standard derivation of the diffusion
equation can be generalized into a ''nonextensive'' form without the
introduction of \emph{ad hoc} modifications. Instead, we simply generalize
the ansatz used to resum the expansion of the intermediate scattering function. 
For an intial condition of a Dirac delta function in real space, the solutions
of the resulting diffusion equation in Fourier space take the form of the $q$%
-exponential which commonly arises in nonextensive statistical mechanics.
This solution corresponds to normal diffusion, but with non-classical
distributions. Therefore the $q$-generating function approach developed here
should be useful to similarly generalize other derivations of diffusive transport. 


\vspace{2cm}

{\bf FIGURE CAPTIONS}

\bigskip

FIGURE 1 : Distribution $G_{q>1}(r,t)$ as a function of $r^{\ast}=r/\protect\sqrt{Dt}$.

\bigskip

FIGURE 2 : Distribution $G_{q}(r,t)$ as a function of $r^{\ast}=r/\protect\sqrt{Dt}$
for $q=1.05$ and $q=0.95$ compared to the Gaussian solution.

\bigskip

FIGURE 3 : Distribution $G_{q<1}(r,t)$ as a function of $r^{\ast}=r/\protect\sqrt{Dt}$.

\newpage

\begin{figure*}[htbp]
\includegraphics[angle=0,scale=0.3]{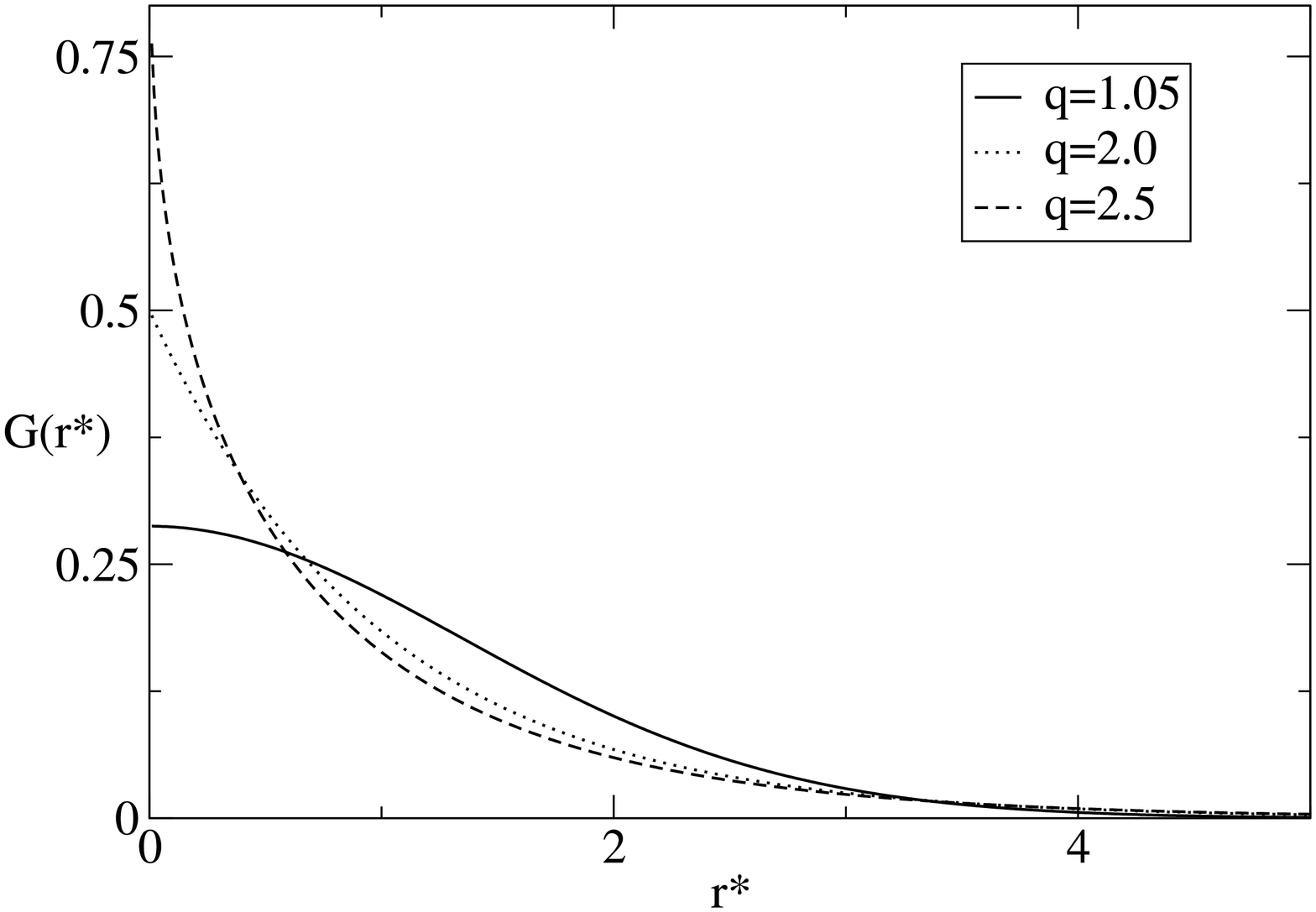}
\label{fig1}
\end{figure*}

\begin{figure*}[htbp]
\includegraphics[angle=0,scale=0.3]{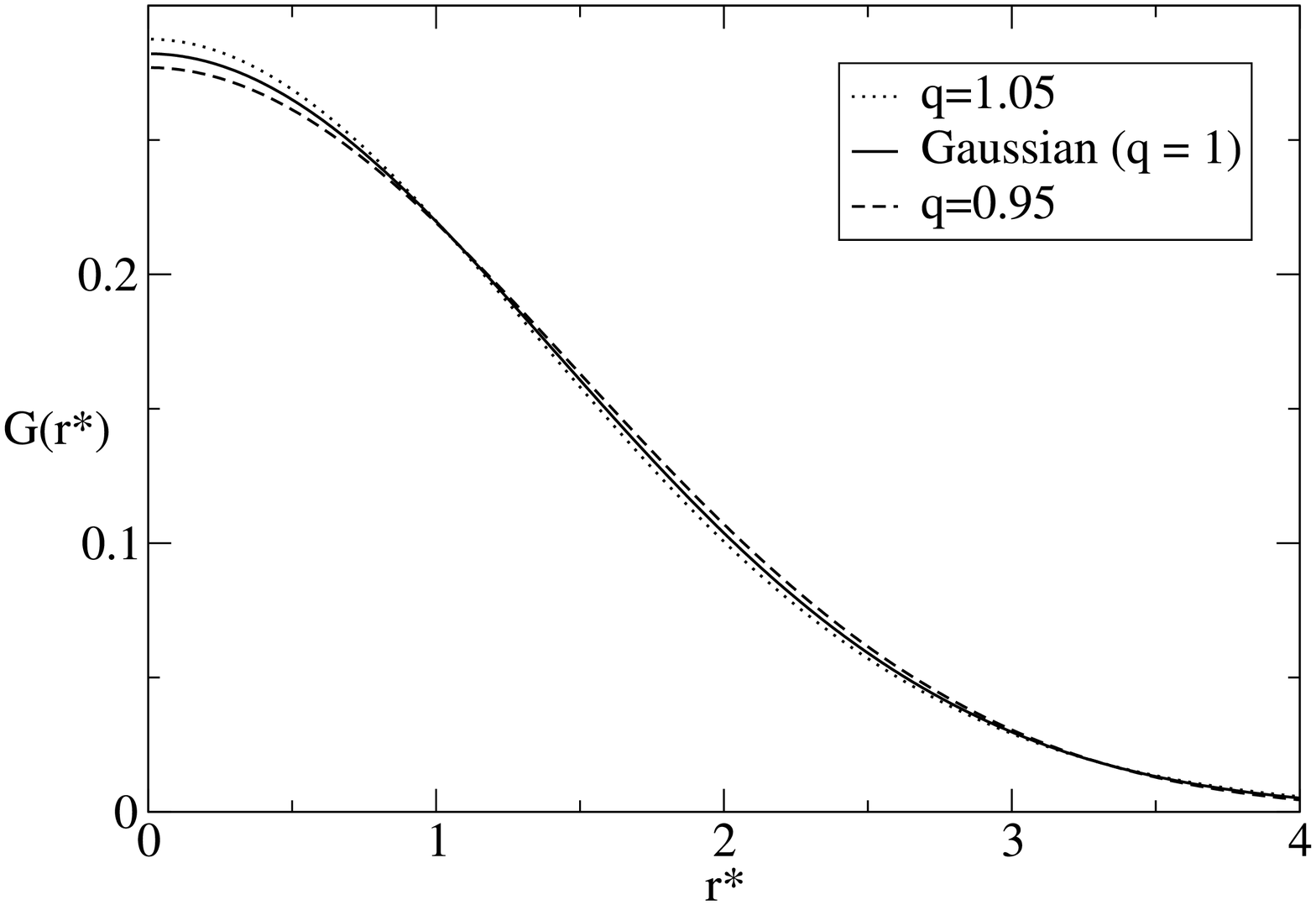}
\label{fig2}
\end{figure*}

\begin{figure*}[htbp]
\includegraphics[angle=0,scale=0.3]{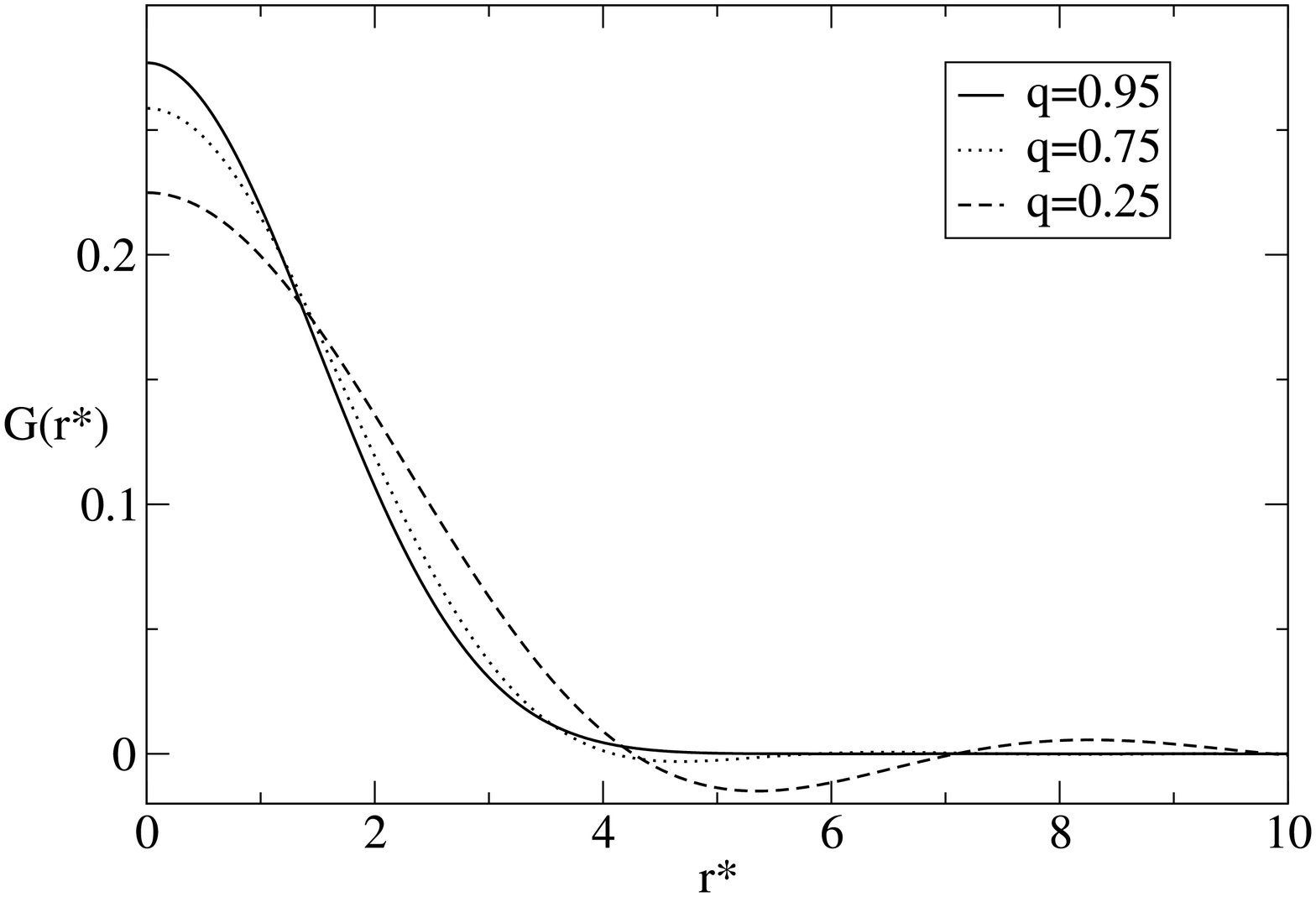}
\label{fig3}
\end{figure*}


\begin{thebibliography}{9}
\bibitem{footnote} Notice that in both cases the generalized diffusion coefficient exhibits
singularities.
\end{thebibliography}
\end{document}